\documentclass[twocolumn,prl,floatfix,superscriptaddress,longbibliography]{revtex4-2}
\usepackage[hidelinks,colorlinks=true,citecolor=blue]{hyperref}

\usepackage{float}
\usepackage[dvipsnames]{xcolor}
\usepackage{graphicx}
\usepackage{amsmath}
\usepackage{bm}
\usepackage{ulem}
\usepackage{stix} 

\begin{document}

\title{Site-polarized Mott phases competing with a correlated metal in twisted WSe$_2$}

\author{Siheon Ryee}
 \email{siheon.ryee@uni-hamburg.de} 
\affiliation{I. Institute of Theoretical Physics, University of Hamburg, Notkestra{\ss}e 9-11, 22607 Hamburg, Germany}
\affiliation{The Hamburg Centre for Ultrafast Imaging, Luruper Chaussee 149, 22761 Hamburg, Germany}

\author{Lennart Klebl}
\affiliation{Institute for Theoretical Physics and Astrophysics
	and Würzburg-Dresden Cluster of Excellence ct.qmat,
	University of Würzburg, 97074 Würzburg, Germany}
\affiliation{I. Institute of Theoretical Physics, University of Hamburg, Notkestra{\ss}e 9-11, 22607 Hamburg, Germany}

\author{Gautam Rai}
\affiliation{I. Institute of Theoretical Physics, University of Hamburg, Notkestra{\ss}e 9-11, 22607 Hamburg, Germany}
\affiliation{The Hamburg Centre for Ultrafast Imaging, Luruper Chaussee 149, 22761 Hamburg, Germany}

\author{Ammon Fischer}
\affiliation{Max Planck Institute for the Structure and Dynamics of Matter, Center for Free Electron Laser Science, 22761 Hamburg, Germany}
\affiliation{Institute for Theory of Statistical Physics, RWTH Aachen University, and JARA Fundamentals of Future Information Technology, 52062 Aachen, Germany}

\author{Valentin Cr\'epel}
\affiliation{Center for Computational Quantum Physics, Flatiron Institute, New York, NY 10010, USA}

\author{Lede Xian}
\affiliation{Tsientang Institute for Advanced Study, Zhejiang 310024, China}
\affiliation{Songshan Lake Materials Laboratory, 523808 Dongguan, Guangdong, China}
\affiliation{Max Planck Institute for the Structure and Dynamics of Matter, Center for Free Electron Laser Science, 22761 Hamburg, Germany}

\author{Angel Rubio}
\affiliation{Max Planck Institute for the Structure and Dynamics of Matter, Center for Free Electron Laser Science, 22761 Hamburg, Germany}
\affiliation{Center for Computational Quantum Physics, Flatiron Institute, New York, NY 10010, USA}

\author{Dante M. Kennes}
\affiliation{Institute for Theory of Statistical Physics, RWTH Aachen University, and JARA Fundamentals of Future Information Technology, 52062 Aachen, Germany}
\affiliation{Max Planck Institute for the Structure and Dynamics of Matter, Center for Free Electron Laser Science, 22761 Hamburg, Germany}

\author{Roser Valent\'i}
\affiliation{Institut für Theoretische Physik, Goethe Universit{\"a}t Frankfurt, Max-von-Laue-Strasse 1, 60438 Frankfurt am Main, Germany}

\author{Andrew J. Millis}
\affiliation{Center for Computational Quantum Physics, Flatiron Institute, New York, NY 10010, USA}
\affiliation{Department of Physics, Columbia University, 538 West 120th Street, New York, NY 10027, USA}

\author{Antoine Georges}
\affiliation{Coll\`ege de France, 11 place Marcelin Berthelot, 75005 Paris, France}
\affiliation{Center for Computational Quantum Physics, Flatiron Institute, New York, NY 10010, USA}
\affiliation{CPHT, CNRS, Ecole Polytechnique, Institut Polytechnique de Paris, Route de Saclay, 91128 Palaiseau, France}
\affiliation{DQMP, Universit{\'e} de Gen{\`e}ve, 24 quai Ernest Ansermet, CH-1211 Gen{\`e}ve, Suisse}

\author{Tim O. Wehling}
\affiliation{I. Institute of Theoretical Physics, University of Hamburg, Notkestra{\ss}e 9-11, 22607 Hamburg, Germany}
\affiliation{The Hamburg Centre for Ultrafast Imaging, Luruper Chaussee 149, 22761 Hamburg, Germany}

\date{\today}

\begin{abstract}
Twisted WSe$_2$ hosts superconductivity, metal-insulator phase transitions, and field-controllable Fermi-liquid to non-Fermi-liquid transport properties. 
In this work, we use dynamical mean-field theory to provide a coherent understanding of the electronic correlations shaping the twisted WSe$_2$ phase diagram. We find a correlated metal competing with three distinct site-polarized correlated insulators; the competition is controlled by interlayer potential difference and interaction strength. The insulators are characterized by a strong differentiation between orbitals with respect to carrier concentration and effective correlation strength. Upon doping, a strong particle-hole asymmetry emerges, resulting from a Zaanen-Sawatzky-Allen-type charge-transfer mechanism. The associated charge-transfer physics and proximity to a van Hove singularity in the correlated metal sandwiched between two site-polarized insulators naturally explains the interlayer potential-driven metal-to-insulator transition, particle-hole asymmetry in transport, and the coherence-incoherence crossover in $3.65^\circ$ twisted WSe$_2$.
\end{abstract}
 
\maketitle

Twisted transition-metal dichalcogenide homobilayers are emerging platforms for exploring correlation phenomena~\cite{cai2023, zeng2023, xu2023, park2023, devakul2021magic, kang2024double, kang2024, crepel2024spinon, ghiotto2024stoner, anderson2023programming, crepel2023anomalous, ghiotto2021quantum, li2021continuous, zang2022dynamical, wei2024, belanger2022, wietek2022tunable, wu2023pair, klebl2023competition, zegrodnik2023, wang2020correlated, ghiotto2021quantum, zhang2021electronic, zang2021, pan2021, zang2022dynamical, ryee2023switching, knuppel2025, tscheppe2024} such as superconductivity~\cite{belanger2022, wietek2022tunable, witt2022, wu2023pair, klebl2023competition, zegrodnik2023} and metal-insulator transitions~\cite{wang2020correlated, ghiotto2021quantum, zhang2021electronic, zang2021, pan2021, zang2022dynamical, ryee2023switching, knuppel2025, tscheppe2024}.
At the heart of this rich landscape of correlated electron physics is an extraordinarily high level of experimental control. The relative strengths of interaction and kinetic energies are controlled by twist angle $\theta$, while important aspects of the band structure including the energy of the van Hove singularity (vHS) may be tuned {\it in situ} by "displacement field" (layer dependent electronic potential) $E_z$  arising from a potential difference between top and bottom gates while carrier concentration per unit cell may also be varied over wide ranges by varying the average of the top and bottom gate potentials~\cite{wu2019topological,pan2020,Kennes2021}.

The recent discovery of superconductivity and neighboring correlated phases in twisted WSe$_2$ (tWSe$_2$) at two distinct twist angles ($\theta = 3.65^\circ$ and $5^\circ$)~\cite{xia2024unconventional,guo2024superconductivity} has sparked a surge of theoretical interest in this direction~\cite{schrade2021nematic,kim2024theory,zhu2024theory,christos2024approximate,guerci2024topologicalSC,tuo2024theorytopo,chubukov2024quantum,xie2024superconductivity,qin2024kohn,chubukov2024quantum,fischer2024theory, xie2025kondo, munoz2025twist}. The less strongly correlated $\theta = 5^\circ$  system exhibits superconductivity in proximity to a Fermi-surface reconstructed (presumably antiferromagnetic) phase, with both phases occurring only in a small range of incommensurate hole fillings at which the Fermi surface is in proximity to the van Hove singularity~\cite{guo2024superconductivity}. 
The more strongly correlated $\theta = 3.65^\circ$ system exhibits superconductivity at carrier concentrations very close to the commensurate hole filling of $\nu = 1$ and at very small displacement fields, transitioning at $\nu=1$ into a correlated insulator upon applying an $E_z$~\cite{xia2024unconventional} that is much smaller than the bandwidth, indicating a strong effect beyond a single-particle description.
$3.65^\circ$ tWSe$_2$ also exhibits unusual transport properties including a maximum at a characteristic temperature $T^* \sim 10$~K in the normal state, and a strong particle-hole asymmetry in the magnitude of longitudinal resistance $R$ and in the Fermi-liquid (FL) coherence temperature $T_\mathrm{FL}$ below which $R \propto T^2$ ($T$ denotes temperature). Although the pairing mechanism of superconductivity is yet to be fully understood, its proximity to such correlated phases and unusual transport  implies a potential connection between them; see e.g.,~Refs.~\cite{guerci2024topologicalSC,fischer2024theory}.

\begin{figure} [!htbp] 
	\includegraphics[width=0.9\columnwidth, angle=0]{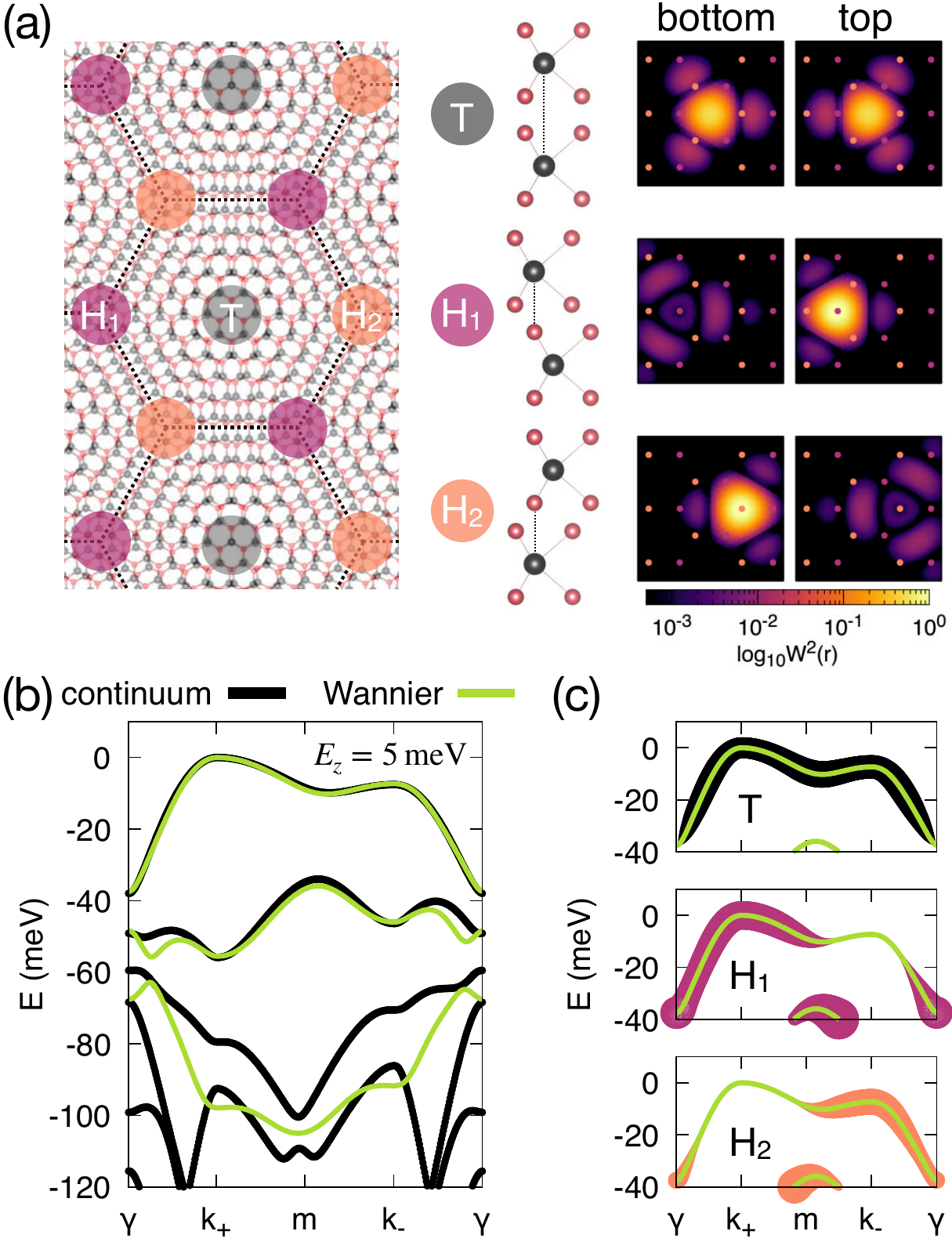}
	\caption{(a) Sketch of the local MM (T), MX (H$_1$), and XM (H$_2$) stacking regions in the moir\'e lattice of tWSe$_2$ and the amplitude of the corresponding Wannier functions on the top layer ($l=+1$) and the bottom layer ($l=-1$) at $E_z=0$. (b) Three-orbital Wannier bands for one valley/spin ($\sigma= \uparrow$) overlaid with continuum model bands at $E_z=5$~meV along with (c) the corresponding spectral weight of Wannier orbitals in the topmost band. The opposite valley/spin ($\sigma= \downarrow$) states are obtained via time-reversal conjugation of the $ \uparrow$ states.
	}
	\label{fig1}
\end{figure}

Here, we propose a physical mechanism underlying the correlated electron behavior observed in $3.65^\circ$ tWSe$_2$ which we validate by applying dynamical mean-field theory (DMFT)~\cite{georges1992,georges1996dynamical} to a multiorbital model that faithfully captures the band structure and topology~\cite{qiu2023,crepel2024bridging,fischer2024theory}. The model includes three localized orbitals per moir\'e unit cell, labeled T, H$_1$, and H$_2$, centered at positions corresponding to local MM, MX, and XM stackings (where M denotes a transition-metal atom and X a chalcogen atom), respectively, as illustrated in Fig.~\ref{fig1}(a). We demonstrate that a Mott MIT occurs at $\nu=1$ induced either by applying $E_z$ or tuning the dielectric constant $\varepsilon$. 
The Mott insulating state features a large polarization of orbital (or site) occupations with one of the T, H$_1$, and H$_2$ sites occupied by approximately one hole and the other two having a very low hole occupancy. The occupancy difference  leads to strong orbital differentiation in effective correlation strength and to a strong particle-hole asymmetry upon doping. The doping asymmetry arises from charge transfer physics similar to that discussed by Zaanen, Sawatzky, and Allen (ZSA)~\cite{ZSA}. Remarkably, the combination of  charge-transfer physics and the proximity to a vHS in a correlated metal sandwiched between two distinct site-polarized insulators naturally explain several key features observed in the experiment.

Our Hamiltonian reads $\mathcal{H} = H_0 + H_\mathrm{int}$. Here, $H_0$ is the kinetic term that accounts for the band structure and includes a layer-dependent potential $l E_z$ (with $l=\pm1$ denoting the two layers). Because the H$_1$ and $H_2$ orbitals are localized on the top and bottom orbitals, respectively, while the wave function of the T orbital is equally divided between the two layers [Fig.~\ref{fig1}(a)], a non-zero displacement field splits the on-site energies of the H$_1$ and H$_2$ orbitals while leaving the on-site energy of the T orbital unchanged.
We construct $H_0$ via a one-shot Wannierization using an energy-dependent weighting factor applied to the entire Bloch states~\cite{marzari2012maximally, carr2019derivation, fischer2024supercell}; see Supplementary~\cite{supple} and Ref.~\cite{fischer2024theory}. 
Figure~\ref{fig1}(b) compares the continuum bands calculated at $E_z=0$ (green lines) to the Wannier band structure resulting from diagonalization of $H_0$ at $E_z=5$~meV (black lines).  The topmost band, which is the most relevant near $\nu=1$, is perfectly reproduced, in both dispersion and band topology (not shown here) and the second band is very well reproduced.
The third band (whose details are not relevant to our calculations) represents an average over the spectral weight of the remote continuum bands, which must be included to have a model with three exponentially localized Wannier orbitals with a vanishing total Chern number~\cite{crepel2024bridging,fischer2024theory}.
 All three orbitals have substantial spectral weight in the topmost band [Fig.~\ref{fig1}(c)], with the H$_1$ (H$_2$) orbital contributing predominantly near the $k_+$ ($k_-$) point, reflecting the nearly complete layer polarization of the corresponding Wannier function [Fig.~\ref{fig1}(a)]. At $\nu = 1$ and $E_z = 0$, the orbital-resolved hole occupations are $\nu_\mathrm{T} \simeq 0.5$ and $\nu_\mathrm{H_1} = \nu_\mathrm{H_2} \simeq 0.25$.

The interaction term $H_\mathrm{int}$ accounts for local and nonlocal density-density Coulomb interactions, which reads
\begin{align} \label{eq_int}
	H_\mathrm{int} = \sum_{i, \eta} \frac{U_\eta}{\varepsilon}n_{i \eta \uparrow}n_{i \eta  \downarrow} + \frac{1}{2} \sum^{i \eta \neq j\eta'}_{ij, \eta \eta', \sigma \sigma'} \frac{ U_{i\eta,j\eta'} } {\varepsilon} n_{i \eta \sigma}n_{j \eta'  \sigma'},
\end{align} 
where $i,j$ denote Bravais lattice cells, $\eta,\eta' \in \{\mathrm{T}, \mathrm{H}_1, \mathrm{H}_2\}$ the orbital, and $\sigma, \sigma' \in \{\uparrow, \downarrow \}$ the spin locked to the valley degree of freedom~\cite{wu2019topological,wang2020correlated}. $n_{i\eta\sigma}$ is the associated hole number operator.
$U_{i\eta, j \eta'}$ refers to Coulomb interaction between orbitals $\eta$ and $\eta'$ located in Bravais lattice cells $i$ and $j$, respectively, obtained from evaluating double-gate Coulomb interaction in the Wannier basis at $E_z=0$. $U_\eta \equiv U_{i\eta,i \eta}$ and $U_\mathrm{H}  \equiv U_{\mathrm{H}_1} = U_{\mathrm{H}_2}$.  We find that $ U_\mathrm{H} \simeq 1.7U_{\mathrm{T}}$ due to the stronger localization of $\mathrm{H}_1$ and $\mathrm{H}_2$ Wannier orbitals compared to that of the $\mathrm{T}$ orbital.
We rescale Coulomb interactions by introducing an adjustable parameter $\varepsilon$, which acts as a dielectric constant; see Supplementary~\cite{supple}. Non-density-density interactions are at least an order of magnitude smaller and are thus neglected.

Our DMFT calculations treat local interactions
[the first term in Eq.~(\ref{eq_int})] with the numerically exact hybridization-expansion continuous-time quantum Monte Carlo method~\cite{werner2006, gull2006, comctqmc}, and nonlocal interactions [the second term in Eq.~(\ref{eq_int})] with the Hartree approximation. 
The Fock self-energy is neglected to prevent bandwidth overestimation \cite{ayral2017}. The model is solved in the hole representation (see Supplementary~\cite{supple}). Since there are three sites per unit cell, three impurity problems must be solved in every DMFT self-consistency loop. Where necessary, the self-energy is analytically continued from Matsubara to real frequency ($\omega$) axis using the maximum entropy method~\cite{jarrel1996,triqs_maxent,supple}.

\begin{figure} [!htbp]	 
	\includegraphics[width=1.0\columnwidth]{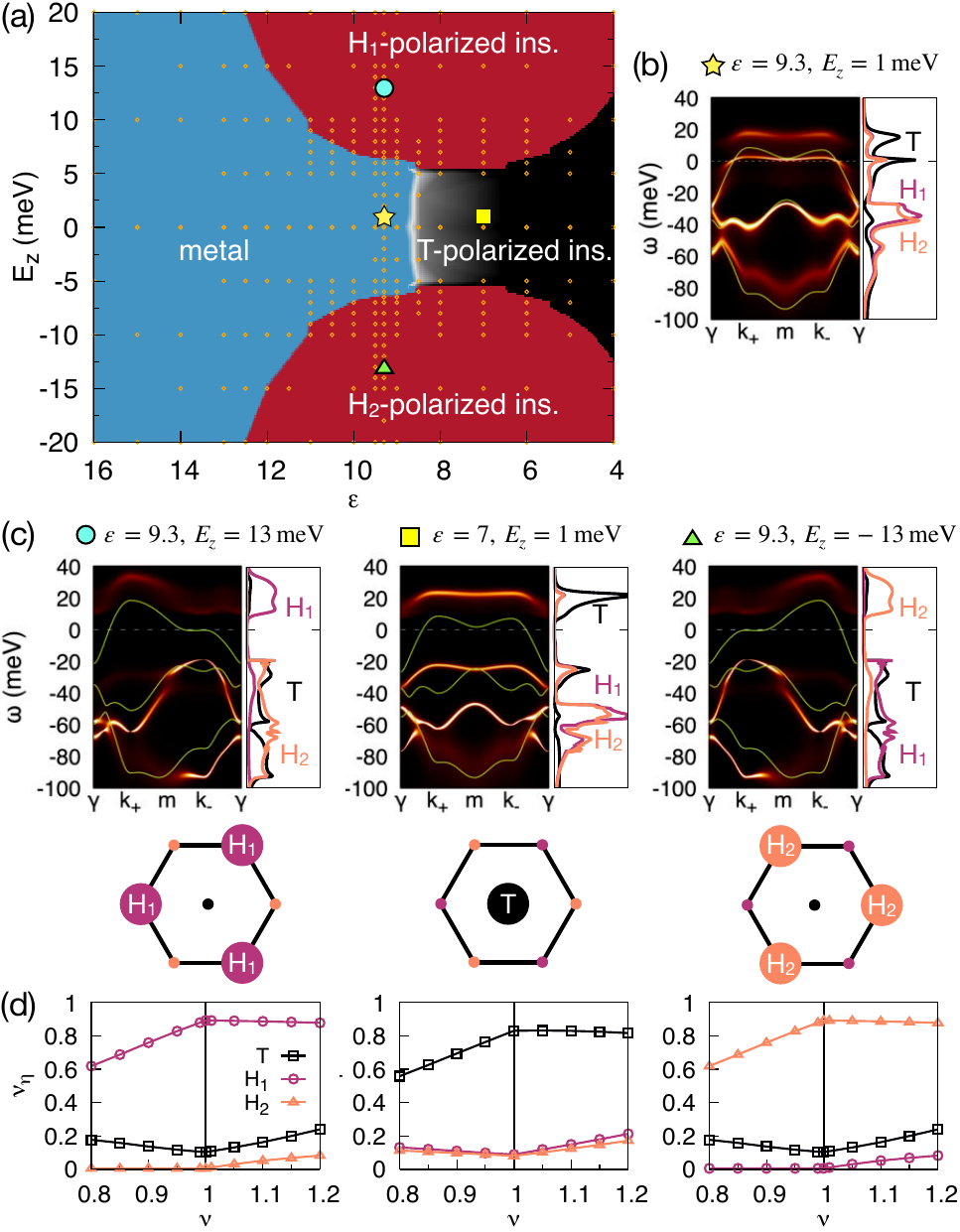}
	\caption{(a) Phase diagram in the $\varepsilon$-$E_z$ plane at hole filling $\nu=1$ and $T=5.8$~K. 
    (b,c) Momentum-resolved and momentum-integrated spectral functions for $\sigma = \uparrow$ and $\nu=1$ obtained at the ($\varepsilon,~E_z$) values marked by colored symbols in the phase diagram. The continuum-model bands (green) are overlaid. The schematics in the bottom panels of (c) illustrate the hole occupation at each site, represented by the size of the circles for the H$_1$-polarized (left), T-polarized (middle), and H$_2$-polarized (right) insulators.  
	(d)  The orbital-resolved hole occupation $\nu_\eta$ as a function of total occupation $\nu$ for the same ($\varepsilon,~E_z$) sets as in the upper panels in (c). The symmetry of the system under $H_1 \rightarrow H_2$ upon $E_z \rightarrow -E_z$ leads to a corresponding symmetry of the spectral functions (c) and of the orbital occupancies (d).
	}
	\label{fig2}
\end{figure}

The DMFT phase diagram in the $\varepsilon$-$E_z$ plane for $\nu=1$ is shown in Fig.~\ref{fig2}(a).
We first identify a correlated metallic phase near $\varepsilon \sim 10$ and in the small $|E_z|$ regime.
It is characterized by dominant T-orbital hole occupation and a small quasiparticle weight $Z$ at $\omega=0$, accompanied by incoherent spectral weight at higher energies [cf.~Fig.~\ref{fig2}(b)]. 
Interestingly, this metal undergoes first-order Mott transitions to three distinct correlated insulating phases upon a slight tuning of either $E_z$ or $\varepsilon$.

The three correlated insulators are characterized by a strong orbital polarization in the hole occupation, which results in localization of holes at the sites hosting the relevant orbital, as depicted in the bottom panels of Fig.~\ref{fig2}(c). This polarization is influenced by $E_z$ which introduces energy difference between the top and bottom layers: $E_z$ shifts the hole energy level of H$_1$ (H$_2$) orbital approximately by $-E_z$ ($+ E_z$). 
Correspondingly, we find H$_1$- and H$_2$-polarized insulators for $E_z>0$ and $E_z<0$ regimes, respectively.
This transition is facilitated by Coulomb interactions, so the phase boundaries between metal and the insulators are shifted toward smaller $|E_z|$ as $\varepsilon$ decreases. This underscores the decisive role of Coulomb interactions and naturally explains how a small $E_z$ can induce the MIT as observed in experiment \cite{xia2024unconventional}. Here, nonlocal interactions between orbitals of different character, as well as the local interactions, promote site-polarized localization by penalizing a homogeneous charge distribution among orbitals.
Below a critical $\varepsilon$, Coulomb interactions for the T orbital also become large enough to localize T-orbital holes, resulting in a Mott transition in the T-orbital sector as shown in Fig.~\ref{fig2}(a). In this regime, the T-polarized insulator preempts the formation of H$_1$- and H$_2$-polarized insulators.

Further insight can be obtained by examining spectral functions shown in Fig.~\ref{fig2}(c). Note that the spectral functions are defined here for electrons, so that increasing the hole occupancy $\nu$ corresponds to populating fewer states as a function of $\omega$.
For the three distinct site-polarized insulators, a Mott gap opens in the nearly half-filled orbital sector, whereas the spectral weight of the remaining orbitals lies below the chemical potential—indicating the absence of hole occupation—in sharp contrast to the noninteracting case [cf. Fig.~\ref{fig1}(c)]. 
This polarization leads to strong site differentiation in effective correlation strength, as defined by the imaginary part of the DMFT self-energy: the nearly half-filled orbital exhibits strong correlations subject to a Mott transition (large self energy), while the nearly empty (i.e., no holes) orbitals are weakly correlated (small self energy). 

A clear dichotomy emerges between electron-doped and hole-doped sides of these insulators near $\nu=1$. As inferred from the spectral functions in Fig.~\ref{fig2}(c), electrons are doped into strongly correlated orbitals, whereas holes are doped into weakly correlated ones. This is corroborated by the orbital-resolved hole occupation $\nu_\eta$ obtained from DMFT around $\nu=1$ presented in Fig.~\ref{fig2}(d). 
Namely, for $\nu > 1$, holes are doped into the weakly correlated orbitals, whereas $\nu_\eta$ for the strongly correlated orbitals remains nearly unchanged.
In this respect, these site-polarized insulators can be classified as charge-transfer insulators in the ZSA scheme~\cite{ZSA}.

\begin{figure*} [!htbp] 
	\includegraphics[width=0.95\textwidth]{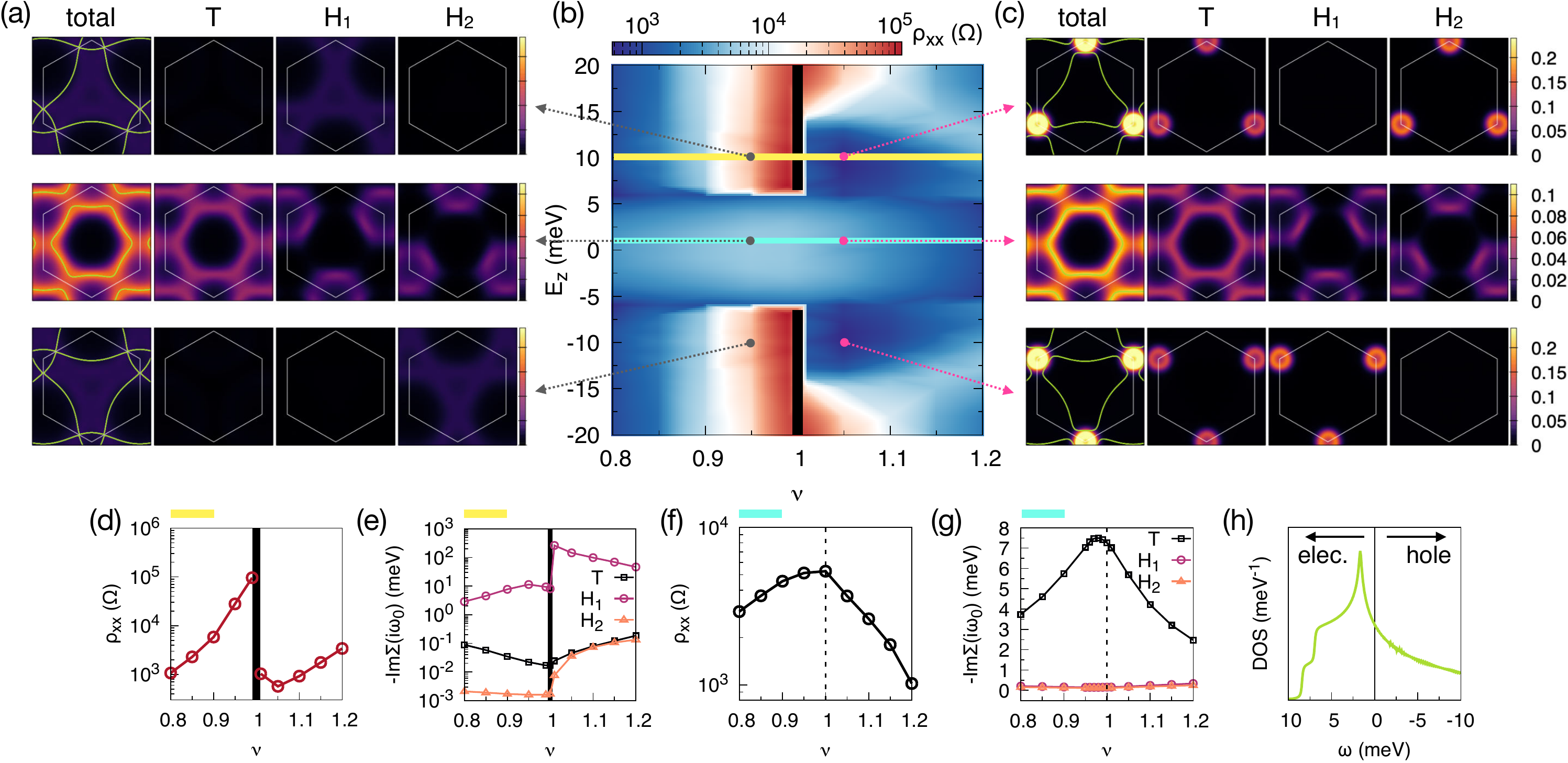}
	\caption{(a,c) $\bm{k}$-resolved spectral functions of spin-$\uparrow$ character at the Fermi level in the Brillouin zone for (a) electron-doped and (c) hole-doped regions, approximated by the DMFT Green's function $-G(\bm{k}, \tau = \frac{1}{2T})/(\pi T)$ with $\tau$ imaginary time. Noninteracting Fermi surface (green line) is overlaid. (b) Calculated resistivity $\rho_{xx}$ using the DMFT self-energy as a function of $\nu$ and $E_z$ at $\varepsilon=9.3$ and $5.8$~K. The black bars indicate the insulating regions. (d,~f) $\rho_{xx}$ along the yellow and cyan lines indicated in (b).
    (e,~g) Scattering rate $\Gamma$ which is approximated by the imaginary part of the self-energy at the lowest Matsubara frequency $\omega_0=\pi T$, $\Gamma = -\mathrm{Im}\Sigma(i\omega_0 = i\pi T)$, calculated along the yellow and cyan lines in (b). (h) Noninteracting density of states at $E_z=1$~meV. The vertical black solid line at $\omega=0$ indicates the Fermi level at $\nu=1$.
	}
	\label{fig3}
\end{figure*}

Motivated by the recent experiment by Xia et al.~\cite{xia2024unconventional}, we next focus on the transport properties. In DMFT, the conductivity can be obtained as~\cite{georges1996dynamical} 
\begin{align} \label{kubo}
	\sigma^\mathrm{dc}_{\alpha \beta} =  \frac{\pi e^2}{ \hbar S N }  \sum_{\sigma \bm{k}}  \int \! d\omega \,\Big({-\frac{\partial f}{\partial \omega}}\Big) \mathrm{Tr}[ \bm{v}^\alpha_{\sigma \bm{k}} \bm{A}_{\sigma \bm{k}}(\omega) \bm{v}^\beta_{\sigma \bm{k}} \bm{A}_{\sigma \bm{k}}(\omega) ] \,,
\end{align}
where $S$ denotes the area of the unit cell, $N$ the number of $\bm{k}$-points in the first Brillouin zone, and $f$ the Fermi-Dirac function. $\bm{v}^\alpha_{\sigma \bm{k}}$ is the Fermi velocity incorporating intra unit cell terms in direction $\alpha \in \{x, y\}$~\cite{tomczak2009} and $\bm{A}_{\sigma \bm{k}}(\omega)$ the momentum-resolved spectral function, both in the orbital basis; see Supplementary~\cite{supple}. Tr means the trace over orbital indices.
Eq.~(\ref{kubo}) neglects vertex corrections. This is valid in DMFT for single-band systems and extends to multi-orbital systems under assumptions discussed in Ref.~~\cite{dang2015}.

Figure~\ref{fig3}(b) shows the resistivity $\rho_{xx} = [\bm{\sigma}^\mathrm{dc} ]^{-1}_{xx}$ around $\nu=1$ at $\varepsilon=9.3$ and 5.8~K. In good agreement with the experiment~\cite{xia2024unconventional}, two strongly resistive regions sandwich the metallic phase at $\nu = 1$, along with a pronounced asymmetry in $\rho_{xx}$, which is larger on the electron-doped side ($\nu < 1$) than on the hole-doped side ($\nu > 1$). As discussed earlier, these two strongly resistive regions correspond to the H$_1$- and H$_2$-polarized insulators for $E_z>0$ and $E_z<0$ regimes, respectively.
The asymmetry in $\rho_{xx}$ for these H$_1$- and H$_2$-polarized insulators originates from the charge-transfer physics, which induces a strong dichotomy in the orbital character of the low-energy spectral weight between the electron-doped and the hole-doped sides. This is clearly manifested in the $\bm{k}$-resolved spectral function at the Fermi level $A_{\bm k}(0)$. On the electron-doped side [$\nu<1$,~Fig.~\ref{fig3}(a)], only the strongly correlated orbital, namely, H$_1$ for $E_z>0$ (top row) or H$_2$ for $E_z<0$ (bottom row), constitutes the spectral weight at the Fermi level. This results in  an incoherent Fermi surface due to a large scattering rate, and thus bad conductivity; see also Fig.~\ref{fig3}(d). On the other hand, on the hole-doped side [$\nu>1$,~Fig.~\ref{fig3}(c)], the strongly correlated orbital is gapped out and holes are doped to weakly correlated orbitals due to the charge-transfer nature of the insulator [cf.~Fig.~\ref{fig2}(d)]. This leads to a coherent Fermi surface and good conductivity, and explains why the electron-doped side shows higher resistivity than the hole-doped side, despite the strongly correlated orbital showing a much larger scattering rate on the latter [cf.~Fig.~\ref{fig3}(e)].
As more holes are doped, however, scattering rates of T and $\mathrm{H}_2$ are gradually enhanced followed by an increase in resistivity; see the $\nu > 1.05$ side of Fig.~\ref{fig3}(d).
Note that charge redistribution among the orbitals (due to the site-polarized Mott transition at $\nu = 1$) makes the Fermi surface substantially different compared to the noninteracting one indicated by the green line in Fig.~\ref{fig3}(c).

Interestingly, we also find an asymmetry of $\rho_{xx}$ in the small $|E_z|$ regime, despite the absence of a site-polarized Mott transition at $\nu = 1$; see Fig.~\ref{fig3}(f). In this regime, the T orbital dominates the spectral weight at the Fermi level over the H orbitals as shown in the middle row of Figs.~\ref{fig3}(a,c), and the DMFT Fermi surface closely resembles the noninteracting one, with some smearing due to the large scattering rate $\Gamma$ ($> T$) of the dominant T orbital, as shown in Fig.~\ref{fig3}(g). In principle, both a weakened precursor of the orbital polarization scenario or a mechanism distinct from that near the insulating regime [Fig.~\ref{fig3}(d)] could be responsible for the resistivity asymmetry.

Regarding the latter, a likely contributor to the asymmetry here is the proximity to the vHS inherent to the noninteracting band structure. 
The noninteracting density of states (DOS) $D(\omega)$ presented in Fig.~\ref{fig3}(h) clearly exhibits the vHS on the electron-doped side. This corresponds to smaller velocities on the electron-doped side, and also suppresses the intersite hybridization.
This effect is reflected in the local hybridization function $\Delta_0(i\omega)$ at low frequencies, obtained from the noninteracting local Green's function $G_0(i\omega)$ as $\mathrm{Im}\Delta_0(i0^+) = {\mathrm{Im}G_0(i0^+)}/[\mathrm{Re}G_0(i0^+)^2 + \mathrm{Im}G_0(i0^+)^2 ] \simeq -1/[\pi D(0)]$, where $D(0) = - \mathrm{Im}G_0(i0^+)/\pi$~\cite{mravlje2011}. Thus, a larger DOS suppresses the effective coupling of an orbital (site) with its neighbors, leading to stronger correlations~\cite{mravlje2011, karp2020, lee2020, wu2020, kugler2020, bramberger2021, lee2021, kim2022}. It is indeed evidenced by the T-orbital scattering-rate asymmetry which exhibits larger values near $\nu=1$ on the electron-doped side than on the hole-doped side as shown in Fig.~\ref{fig3}(g). This effect also prevails in a single-orbital model using the topmost band only (see Supplementary~\cite{supple}).

The experiment of Ref.~\cite{xia2024unconventional} shows a re-entrant metallic phase at $\nu=1$ in the regime of very large $|E_z|$ that our results do not capture. There can be different reasons for this: as one possibility, in the very large-$E_z$ regime, the two layers can be effectively decoupled such that the moir\'e potential becomes weakened and the ansatz of the three Wannier functions sketched in Fig. \ref{fig1}(a) remains no longer valid. As a second possibility an interplay of enhanced screening, ordering and non-local correlations could be at play. Indeed, both  fRG~\cite{fischer2024theory} and Hartree-Fock~\cite{munoz2025twist} studies of the same model albeit assuming enhanced dielectric screening find ordering tendencies at small to intermediate $E_z$ that are suppressed at very large $E_z$.

\begin{figure} [!htbp]	 
	\includegraphics[width=0.9\columnwidth]{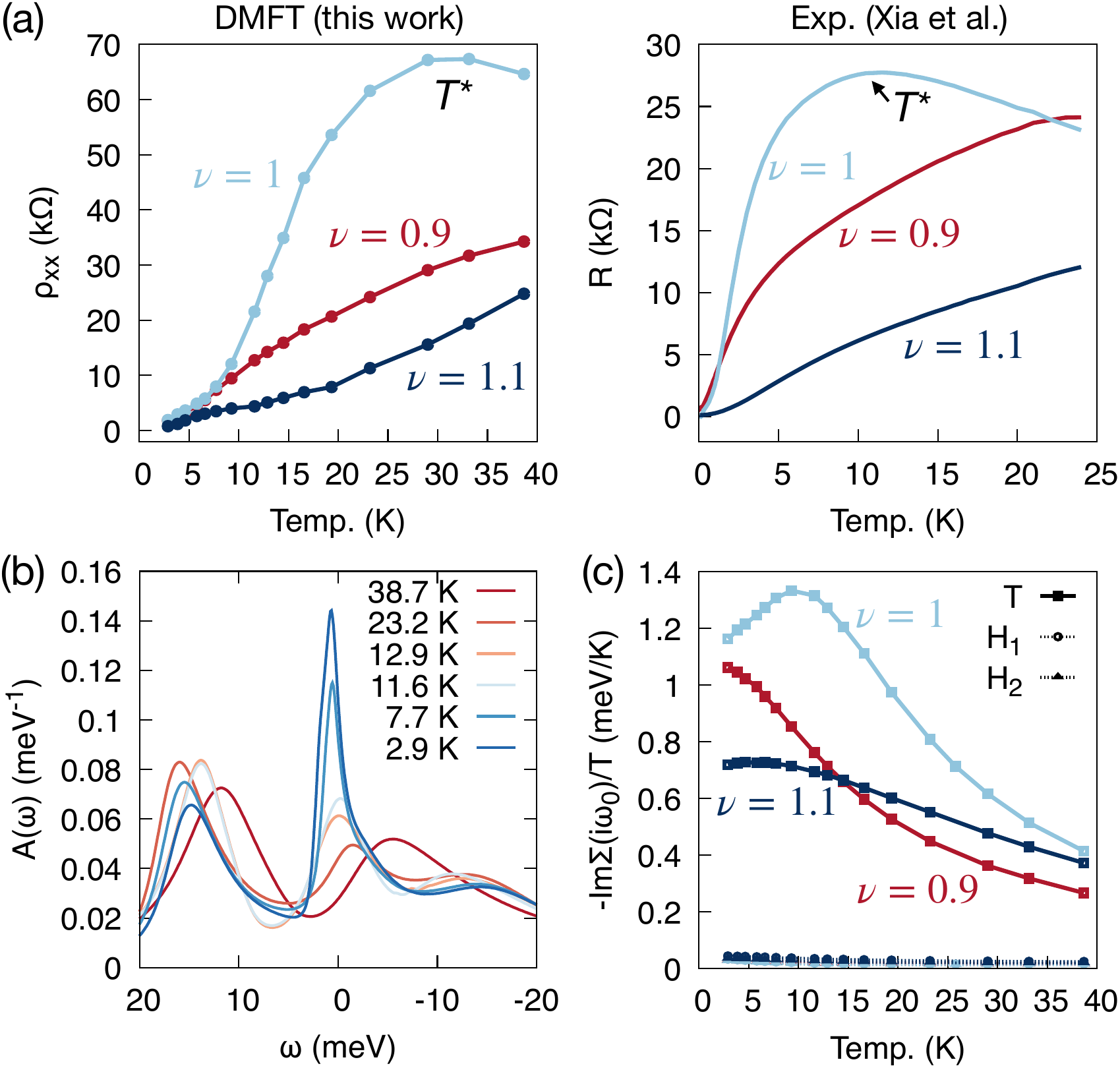}
	\caption{(a) Left: $T$ dependence of the resistivity $\rho_{xx}$ at $\varepsilon=9.3$ and $E_z=1$~meV. Right: The experimental resistance $R$ reported by Xia et al.~at the displacement field of $E \simeq 8~\mathrm{mV}/\mathrm{nm}$~\cite{xia2024unconventional}. (b) Momentum-integrated spectral function $A(\omega)$ for $\nu=1$ at various temperatures. (c) $T$ dependence of  $-\mathrm{Im}\Sigma(i\omega_0)/T$ for the data underlying panel (a).}
	\label{fig4}
\end{figure}

We now investigate the temperature dependence of $\rho_{xx}$ in the metallic phase. In the experiment by Xia et al.~\cite{xia2024unconventional}, superconductivity emerges within a narrow filling range around $\nu = 1$  in the metallic phase with an onset temperature of $\sim 200$~mK. At higher temperatures in the normal state, a characteristic temperature scale $T^* \sim 10$~K is identified, at which the resistance $R$ exhibits a maximum; see the right panel of Fig.~\ref{fig4}(a).  
Our calculated $\rho_{xx}$ in Fig.~\ref{fig4}(a) agrees well with the experimental data. Notably, $\rho_{xx}$ is smallest at $\nu=1.1$, and $\rho_{xx}$ at $\nu=0.9$ and $\nu=1$ become comparable in the low-temperature regime ($T < 10$~K). Additionally, $\rho_{xx}$ exhibits a maximum at $\nu = 1$ as in the experiment, although the corresponding $T^*$ differs by about a factor of three.

The $T^*$ is associated with the destruction of quasiparticles as shown in Fig.~\ref{fig4}(b), which is characteristic of a correlated metal in proximity to a Mott insulator at elevated temperatures~\cite{limelette_2003,deng2013, deng2019, stadler2021}. At very low temperatures, a (resilient) quasiparticle emerges, and thus a sharp peak in the spectral function at $\omega=0$ is identified along with Hubbard bands at higher energies. As the temperature increases, the resilient quasiparticles become progressively shorter-lived, and are completely destroyed in the temperature range between 23.2 and 38.7~K, revealing a correlation-induced  minimum in the density of states resulting from the combination of proximity to a Mott state and the absence of coherent quasiparticles. Thus, the state at $T>T^*$ may be considered to have a pseudogap. We therefore suggest that the resistivity maximum observed in experiment is interpreted as a signature of an incoherent nearly Mott state.

On the doped sides, the pronounced asymmetry between $\nu=0.9$ and $\nu=1.1$ due to the vHS persists up to higher temperatures in agreement with the experiment~\cite{xia2024unconventional}; see Fig.~\ref{fig4}(a). An interesting experimental observation is that the FL coherence temperature $T_\mathrm{FL}$ is significantly larger on the hole-doped side, and rapidly decreases as $\nu$ approaches $\nu=1$ from above ($\nu > 1$). Extracting the $T$ scaling of $\rho_{xx}$ at low temperatures, however, is challenging in our calculations as it requires highly precise analytic continuation of three independent self-energy data (one for each orbital) at every $T$. Instead, we investigate the imaginary part of the self-energy at the lowest Matsubara frequency $\mathrm{Im}\Sigma(i\omega_0)$, as shown in Fig.~\ref{fig4}(c). In a FL regime at sufficiently low temperatures,  $-\mathrm{Im}\Sigma(i\omega_0)/T = (Z^{-1}-1)\pi + \mathcal{O}(T^2)$~\cite{chubukov2012, cha2020, zang2022dynamical}. Thus, $T_{\mathrm{FL}}$ can be identified as the temperature below which $-\mathrm{Im}\Sigma(i\omega_0)/T$ forms a plateau.  Indeed, we find the asymmtery in $T_\mathrm{FL}$ as in the experiment. Namely, $-\mathrm{Im}\Sigma(i\omega_0)/T$ saturates below $\sim 7$~K at $\nu = 1.1$, which is comparable to the experimental $T_\mathrm{FL}$ of 3.5~K~\cite{xia2024unconventional}. In contrast, it continues to increase at $\nu = 0.9$ and decrease at $\nu = 1$ down to the lowest temperature we examined (2.8~K), indicating that $T_\mathrm{FL}$ is lower in those cases. Thus, the vHS here not only increases the magnitude of resistivity, but also suppresses the FL coherence by enhancing electron correlations.

To conclude, we have demonstrated that the charge-transfer physics and proximity to the vHS of the correlated metal sandwiched between two distinct site-polarized insulators provide a coherent explanation for several key features observed in the experiment on $3.65^\circ$ tWSe$_2$~\cite{xia2024unconventional}, including metal-insulator transitions, their tunability with displacement field, doping and temperature dependence of the electronic transport properties, and a coherence-incoherence crossover. 
A central open question regards ordering phenomena, particularly possible spin- or valley-polarized electronic orders emerging from site-polarized Mott insulators and the mechanism by which superconductivity arises in the correlated metallic phase at small $|E_z|$. Understanding how electronic phases evolve with decreasing correlation strength as $\theta\to 5^\circ$ ~\cite{guo2024superconductivity}, and developing a unified framework for superconductivity in tWSe$_2$ remain outstanding challenges. The present study, focused on the strongly correlated regime at $3.65^\circ$, provides a natural starting point and theoretical foundation for this effort.

{\it Acknowledgments.} We thank Kin Fai Mak and Jie Shan for valuable discussions and sharing of data and D. Munoz-Segovia for insightful discussions of the theory.
SR and TOW acknowledge funding and support by the DFG research unit FOR 5242 (WE 5342/7-1, project No. 449119662). 
SR, GR, and TOW acknowledge funding and support from the Cluster of Excellence ``CUI: Advanced Imaging of Matter'' of the DFG (EXC 2056, Project ID 390715994).
GR and TOW acknowledge funding and support from the European Commission via the Graphene Flagship Core Project 3 (Grant Agreement No. 881603) and the DFG priority program SPP 2244 (Project No. 443274199).
LK acknowledges support from the DFG through Project-ID 258499086 -- SFB 1170 and through the W\"urzburg-Dresden Cluster of Excellence on Complexity and Topology in Quantum Matter -- ct.qmat, Project-ID 390858490 -- EXC 2147.
LK, RV, and TOW greatfully acknowledge support from the DFG through FOR 5249 (QUAST, Project No. 449872909).
AF and DMK acknowledge funding by the DFG  within the Priority Program SPP 2244 ``2DMP'' -- 443274199.
LX acknowledges supported by the National Key Research and Development Program of China (Grant No. 2022YFA1403501), Guangdong Basic and Applied Basic Research Foundation (Grant No. 2022B1515120020), the National Natural Science Foundation of China (Grant No. 62341404), Hangzhou Tsientang Education Foundation and the Max Planck Partner group programme.
AR acknowledges support by the European Research Council (ERC-2015-AdG694097), the Cluster of Excellence ‘Advanced Imaging of Matter' (AIM), Grupos Consolidados (IT1249-19) and Deutsche Forschungsgemeinschaft (DFG) -- SFB-925 -- project 170620586.
RV was supported in part by the National Science Foundation under Grants No. NSF PHY-1748958 and No. PHY-2309135.
AJM acknowledges support from Programmable Quantum Materials, an Energy Frontier Research Center funded by the U.S. Department of Energy (DOE), Office of Science, Basic Energy Sciences (BES), under award (DE-SC0019443).
DMK and AR acknowledge support by the Max Planck-New York City Center for Nonequilibrium Quantum Phenomena. 
The Flatiron Institute is a division of the Simons Foundation.


%

\end{document}


\renewcommand{\thepage}{S\arabic{page}}  
\renewcommand{\thesection}{SM\arabic{section}}   
\renewcommand{\thetable}{S\arabic{table}}   
\renewcommand{\thefigure}{S\arabic{figure}}
\renewcommand{\theequation}{S\arabic{equation}}

\renewcommand{\citenumfont}[1]{S#1}
\renewcommand{\bibnumfmt}[1]{[S#1]}

\def\tcb{\textcolor{blue}}
\def\tcr{\textcolor{red}}
\def\tcg{\textcolor{green}}
\def\tcc{\textcolor{cyan}}

\onecolumngrid

\title{Supplemental Material for \\ ``Site-polarized Mott phases competing with a correlated metal in twisted WSe$_2$''}

\author{Siheon Ryee}
\affiliation{I. Institute of Theoretical Physics, University of Hamburg, Notkestra{\ss}e 9-11, 22607 Hamburg, Germany}
\affiliation{The Hamburg Centre for Ultrafast Imaging, Luruper Chaussee 149, 22761 Hamburg, Germany}

\author{Lennart Klebl}
\affiliation{Institute for Theoretical Physics and Astrophysics
	and Würzburg-Dresden Cluster of Excellence ct.qmat,
	University of Würzburg, 97074 Würzburg, Germany}
\affiliation{I. Institute of Theoretical Physics, University of Hamburg, Notkestra{\ss}e 9-11, 22607 Hamburg, Germany}

\author{Gautam Rai}
\affiliation{I. Institute of Theoretical Physics, University of Hamburg, Notkestra{\ss}e 9-11, 22607 Hamburg, Germany}
\affiliation{The Hamburg Centre for Ultrafast Imaging, Luruper Chaussee 149, 22761 Hamburg, Germany}

\author{Ammon Fischer}
\affiliation{Max Planck Institute for the Structure and Dynamics of Matter, Center for Free Electron Laser Science, 22761 Hamburg, Germany}
\affiliation{Institute for Theory of Statistical Physics, RWTH Aachen University, and JARA Fundamentals of Future Information Technology, 52062 Aachen, Germany}

\author{Valentin Cr\'epel}
\affiliation{Center for Computational Quantum Physics, Flatiron Institute, New York, NY 10010, USA}

\author{Lede Xian}
\affiliation{Tsientang Institute for Advanced Study, Zhejiang 310024, China}
\affiliation{Songshan Lake Materials Laboratory, 523808 Dongguan, Guangdong, China}
\affiliation{Max Planck Institute for the Structure and Dynamics of Matter, Center for Free Electron Laser Science, 22761 Hamburg, Germany}

\author{Angel Rubio}
\affiliation{Max Planck Institute for the Structure and Dynamics of Matter, Center for Free Electron Laser Science, 22761 Hamburg, Germany}
\affiliation{Center for Computational Quantum Physics, Flatiron Institute, New York, NY 10010, USA}

\author{Dante M. Kennes}
\affiliation{Institute for Theory of Statistical Physics, RWTH Aachen University, and JARA Fundamentals of Future Information Technology, 52062 Aachen, Germany}
\affiliation{Max Planck Institute for the Structure and Dynamics of Matter, Center for Free Electron Laser Science, 22761 Hamburg, Germany}

\author{Roser Valent\'i}
\affiliation{Institut für Theoretische Physik, Goethe Universit{\"a}t Frankfurt, Max-von-Laue-Strasse 1, 60438 Frankfurt am Main, Germany}

\author{Andrew J. Millis}
\affiliation{Center for Computational Quantum Physics, Flatiron Institute, New York, NY 10010, USA}
\affiliation{Department of Physics, Columbia University, 538 West 120th Street, New York, NY 10027, USA}

\author{Antoine Georges}
\affiliation{Coll\`ege de France, 11 place Marcelin Berthelot, 75005 Paris, France}
\affiliation{Center for Computational Quantum Physics, Flatiron Institute, New York, NY 10010, USA}
\affiliation{CPHT, CNRS, Ecole Polytechnique, Institut Polytechnique de Paris, Route de Saclay, 91128 Palaiseau, France}
\affiliation{DQMP, Universit{\'e} de Gen{\`e}ve, 24 quai Ernest Ansermet, CH-1211 Gen{\`e}ve, Suisse}

\author{Tim O. Wehling}
\affiliation{I. Institute of Theoretical Physics, University of Hamburg, Notkestra{\ss}e 9-11, 22607 Hamburg, Germany}
\affiliation{The Hamburg Centre for Ultrafast Imaging, Luruper Chaussee 149, 22761 Hamburg, Germany}


\maketitle
\tableofcontents
\hypersetup{linkcolor=red}

\newpage

\section{Continuum model}
The continuum model Hamiltonian for spin-$\sigma$ ($\sigma \in \{+,- \}$) moir\'e bands of tWSe$_2$ reads \cite{wu2019topological, devakul2021magic}
\begin{align}
	H^{\sigma} = \begin{pmatrix}
		\frac{-\hbar^2(\mathbf{k}-\sigma k_+)^2}{2m^*} + \Delta_+(\mathbf{r}) & \Delta_{T,\sigma}(\mathbf{r}) \\
		\Delta^\dagger_{T,\sigma}(\mathbf{r}) & \frac{-\hbar^2(\mathbf{k}-\sigma k_-)^2}{2m^*} + \Delta_-(\mathbf{r})
	\end{pmatrix},
	\label{eqS1}
\end{align}
where $k_{\pm}$ are the corners of the moir\'e Brillouin zone, resulting from $\pm\theta/2$ rotation of top ($+$) and bottom ($-$) layers. The intralayer potential $\Delta_{\pm}(\mathbf{r})$ and interlayer tunneling $\Delta_T(\mathbf{r})$ are given by 
\begin{align}
	\begin{split}
		\Delta_{\pm}(\mathbf{r}) &= 2V \sum_{j=1,3,5}\mathrm{cos}(\mathbf{g}_j\cdot \mathbf{r} \pm \psi) \pm E_z , \\
		\Delta_{T}(\mathbf{r}) &= \omega(1+ e^{-i\sigma\mathbf{g}_2 \cdot \mathbf{r}} + e^{-i\sigma\mathbf{g}_3 \cdot \mathbf{r}} ).
	\end{split}
	\label{eqS2}
\end{align}
Here, $\mathbf{g}_j$ are obtained by $(j-1)\pi/3$ counterclockwise rotation of $\mathbf{g}_1 = 4\pi \theta/(a_0\sqrt{3})\hat{x}$ with $a_0$ being the lattice constant of monolayer WSe$_2$ ($a_0=3.32$~\AA). We adopt $m^* = 0.39m_e$ ($m_e$: free electron mass) and $(V,\omega,\psi) = (9~\mathrm{meV},  18~\mathrm{meV}, 128^\circ)$ \cite{devakul2021magic}.

\section{Wannierization}
As outlined in the main text, the continuum model is Wannierized following the procedure outlined in (the supplementary information of) Ref.~\cite{fischer2024theory}. We slightly alter the technical parameters when comparing to Ref.~\cite{fischer2024theory} since the smaller twist angle implies smaller kinetic energy scales. After setting up the continuum Hamiltonian \cref{eqS1} on a regular $24\times24$ grid in the moir\'e Brillouin zone (with a spherical cutoff $\bvec g \leq 3 \bvec g_1$), we perform an eigendecomposition to obtain energies $\epsilon_b^\sigma(\bvec k)$ and Bloch states $u_{o,b}^\sigma(\bvec k)$:
\begin{equation}
    H^\sigma_{o',o}(\bvec k) u_{o,b}^\sigma(\bvec k) = \epsilon_b^\sigma(\bvec k) u_{o',b}^\sigma(\bvec k) \,.
\end{equation}
The ``orbital''{} degree of freedom of the eigenstates ($o$) encompasses both layer $l$ and reciprocal moir\'e lattice vector $\bvec g$. For Wannierization, we Fourier transform the $\bvec g$ dependency of the eigenstates to an intracell real-space dependency; a regular $9\times 9$ mesh within the rhombus-shaped moir\'e unit cell. We then project the Bloch states to the three Gaussian trial orbitals (triangular, honeycomb) following Refs.~\cite{qiu2023, crepel2024bridging, fischer2024theory}, with a real-space spread of $0.2|\bvec a_M|$:
\begin{equation}
    \label{eq:proj}
    \tilde P_{\eta,b}(\bvec k) = W_{\eta, \tilde o}(\bvec k) u^*_{\tilde o, b}(\bvec k) \,,
\end{equation}
where we dropped the spin index $\sigma$ for clarity. Note that the Fourier-transformed orbital index is indicated by $\tilde o$, and $\eta$ labels the Wannier orbital (T, H\textsubscript{1,2}), and $W$ denotes the trial function. Before Löwdin orthonormalization of the projectors \cref{eq:proj}, we multiply them with exponential weighting factors in order to select states close to the valence band edge:
\begin{equation}
    P_{\eta,b}(\bvec k) = e^{-|\epsilon_b(\bvec k)|/\gamma} \tilde P_{\eta,b}(\bvec k) \,,
\end{equation}
where we shift the spectrum such that the valence band edge is at zero energy, i.e., $\max_{b,\bvec k}\epsilon_b(\bvec k) = 0$. As exponential cutoff distance, we use $\gamma = 30\,\mathrm{meV}$. We then perform an SVD on $\hat P(\bvec k)$ to obtain exponentially localized, orthonormal Wannier functions $\mathcal W_{\eta,\tilde o}(\bvec k)$ that perfectly capture the topology of the topmost band:
\begin{equation}
    \hat P(\bvec k) = \hat U(\bvec k) \hat \Sigma(\bvec k) \hat V(\bvec k) \,, \qquad
    \hat{\mathcal W}(\bvec k) = \hat U(\bvec k) \hat V^\dagger(\bvec k) \hat u^T(\bvec k) \,.
\end{equation}

\begin{figure} [!htbp]	 
	\includegraphics[width=0.5\columnwidth]{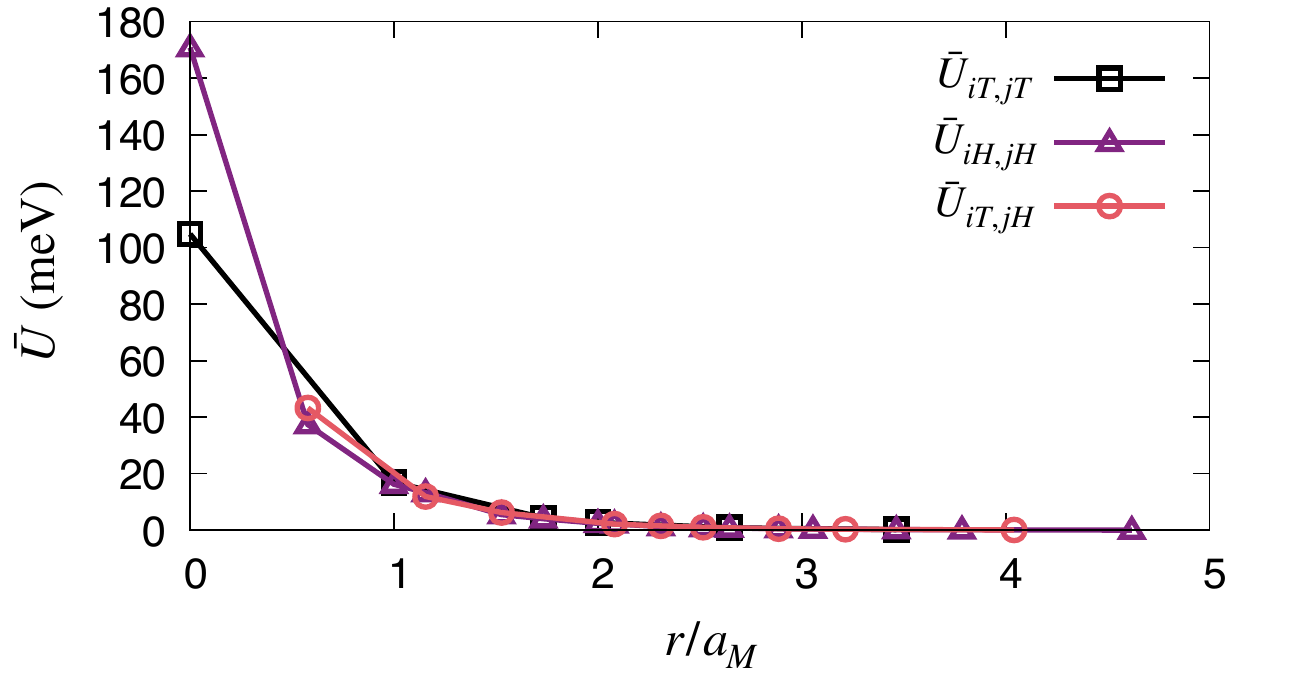}
	\caption{$\bar{U}_{i\eta,j\eta'}$ ($\eta,\eta' \in \{ \mathrm{T},\mathrm{H_1},\mathrm{H_2} \}$) as a function of $r/a_M$, where $a_M$ denotes the lattice constant of the moir\'e unit cell.}
	\label{sfig_U}
\end{figure}

We obtain Coulomb interaction tensor $\bar{\bm{U}}$ by projecting the following short-range regularized dual-gate screened Coulomb interaction~\cite{fischer2024supercell, fischer2024theory, interactpolate} to the real-space Wannier orbital basis:
\begin{equation}
    V(r) = 4V_0 \sum_{k=0}^\infty K_0 \Bigg( \frac{(2k+1)\pi \sqrt{a^2+r^2}}{\xi} \Bigg) \,.
\end{equation}
Here $K_0$ denotes a modified Bessel function of the second kind, $\xi=10\,\mathrm{nm}$ is the distance to the gates, and $a$ and $V_0$ are related to physical parameters through
\begin{equation}
    a = \frac{\alpha}{\varepsilon_0 U} \,, \qquad V_0 = \frac{a}{\varepsilon_0\xi} \,,
\end{equation}
with $U=5\,\mathrm{eV}$ the ``on-site''{} interaction strength in the $9\times 9$ real-space basis and $\varepsilon_0=8$ the dielectric constant. We then evaluated $U_{i\eta,j\eta'}/\varepsilon$ in Eq.~(1) of the main text for arbitrary $\varepsilon$ as 
\begin{equation}
    U_{i\eta,j\eta'}/\varepsilon = (\varepsilon_0/\varepsilon) \bar{U}_{i\eta,j\eta'}
\end{equation}
where $\bar{U}_{i\eta,j\eta'}$ denotes the density-density elements of $\bar{\bm{U}}$ [cf. Fig.~\ref{sfig_U}].

\section{Particle-hole transformation}
Our Hamiltonian is given by $\mathcal{H} = H_0 + H_\mathrm{int} - H_\mathrm{dc}$ in the electron representation. Here $H_0$ is the kinetic term, $H_\mathrm{int}$ the local and nonlocal density-density interaction term, and $H_\mathrm{dc}$ the double-counting term which removes the self-energy included in $H_0$ (which is fitted to the DFT band structure), thereby preventing double counting of the self-energy when solving $\mathcal{H}$ using DMFT. We assume that $H_\mathrm{dc}$ corresponds to a mean-field (Hartree) treatment of $H_\mathrm{int}$ following the spirit of Ref.~\cite{anisimov1991}.
Each term is given explicitly as
\begin{align}
H_0 & =  \sum_{ij, \eta \eta', \sigma}t_{\sigma, i\eta j\eta'} c^{\dagger}_{i\eta \sigma} c_{j\eta' \sigma}, \\
H_\mathrm{int} & = \sum_{i, \eta} \frac{U_\eta}{\varepsilon}\bar{n}_{i \eta \uparrow}\bar{n}_{i \eta  \downarrow} + \frac{1}{2} \sum^{i \eta \neq j\eta'}_{ij, \eta \eta', \sigma \sigma'} \frac{ U_{i\eta,j\eta'} } {\varepsilon} \bar{n}_{i \eta \sigma}\bar{n}_{j \eta'  \sigma'}, \\
H_\mathrm{dc} & = \sum_{i, \eta} \frac{U_\eta}{\varepsilon} \big( \bar{n}_{i \eta \uparrow} \langle \bar{n}_{i \eta  \downarrow} \rangle_0 +  \langle \bar{n}_{i \eta \uparrow}  \rangle_0  \bar{n}_{i \eta  \downarrow}  \big) + \frac{1}{2} \sum^{i \eta \neq j\eta'}_{ij, \eta \eta', \sigma \sigma'} \frac{ U_{i\eta,j\eta'} } {\varepsilon}  \big(  \bar{n}_{i \eta \sigma} \langle \bar{n}_{j \eta'  \sigma'} \rangle_0 +  \langle \bar{n}_{i \eta \sigma} \rangle_0 \bar{n}_{j \eta'  \sigma'} \big),
\end{align} 
where $c_{i \eta \sigma}$ ($c^\dagger_{i \eta \sigma}$) is the electron annihilation (creation) operator for site $i$, orbital $\eta$, and spin $\sigma$. $ \bar{n}_{i\eta\sigma} = c^\dagger_{i \eta \sigma} c_{i \eta \sigma} $ is the corresponding number operator. $t_{\sigma, i\eta j\eta'}$ denotes the hopping amplitudes obtained from the Wannierization of the continuum band structure. $\langle \bar{n}_{i\eta\sigma} \rangle_0$ is the orbital occupation of the original continuum bands. Since we have constructed our model for the case where all continuum bands are occupied, $\langle \bar{n}_{i\eta\sigma} \rangle_0 = 1$.  

We now perform the following particle-hole transformation: 
\begin{align}
c_{i \eta \sigma} \to d^\dagger_{i \eta \sigma},\; c^\dagger_{i \eta \sigma} \to d_{i \eta \sigma}.
\end{align}
Accordingly, the electron number operator transforms as
\begin{align}
\bar{n}_{i\eta\sigma} \to 1 - n_{i\eta\sigma}
\end{align}
where $n_{i\eta\sigma} = d^\dagger_{i \eta \sigma} d_{i \eta \sigma} $ is the hole number operator. Applying this transformation, our Hamiltonian $\mathcal{H}$ in the hole representation reads
\begin{align}
\mathcal{H} = \sum_{ij, \eta \eta', \sigma}-t_{\sigma, i\eta j\eta'} d^{\dagger}_{i\eta \sigma} d_{j\eta' \sigma} + \sum_{i, \eta} \frac{U_\eta}{\varepsilon} n_{i \eta \uparrow} n_{i \eta  \downarrow} + \frac{1}{2} \sum^{i \eta \neq j\eta'}_{ij, \eta \eta', \sigma \sigma'} \frac{ U_{i\eta,j\eta'} } {\varepsilon} n_{i \eta \sigma} n_{j \eta'  \sigma'},
\end{align}
where the first term is the kinetic part $H_0$ in the hole representation and the remaining two terms correspond to Eq.~(1) of our main text, which accounts for interactions between holes.

\section{Analytic continuation of the Matsubara-axis self-energy}

We solve the impurity problem in our DMFT approach with the hybridization-expansion continuous-time quantum Monte Carlo (Hyb-CTQMC) solver, which means that the resulting Green functions and self-energies are on the Matsubara axis. 
In order to extract the $\bm{k}$-resolved spectral function $A(\bm{k}, \omega)$,  we analytically continue the self-energy from the Matsubara axis to the real-frequency ($\omega$) axis. 
This is an ill-posed procedure, meaning that small changes in the input (such as may be induced by QMC noise) can have an out sized effect on the output. 
To regularize the continuation procedure, we use the maximum entropy method as implemented in the triqs\_maxent package~\cite{triqs_maxent}.

In order to analytically continue a Matsubara self-energy $\Sigma(i\omega_n)$ to the real axis, we first generate the corresponding auxiliary Green function $G_\mathrm{aux}(i\omega_n)$. 
We apply the maxent procedure to obtain the auxiliary Green function on the real axis $G_\mathrm{aux}(\omega)$. 
Finally, we reconstruct the real-axis self-energy $\Sigma(\omega)$ from the real-axis auxiliary Green function. 
We define the auxiliary Green function as $G_\mathrm{aux}(z) = \Sigma(z) - \Sigma(i\infty)$ using triqs\_maxent's {\tt{DirectSigmaContinuator}} class. 
Here, $\Sigma(i\infty)$ is the constant term in the high-frequency expansion of $\Sigma(i\omega_n)$. 
The output real-frequency self-energy is sensitive to several tuning parameters of the procedure and care must be taken to avoid artifacts. 
Most of these can be fixed by informed trial and error, making sure that the results are converged.
In the following, we describe the parameter set we used in this study:
\begin{itemize}
    \item We find that the energy window on which the real-frequency auxiliary Green function is defined needs to be substantially larger than the interval of interest  (up to $\pm10$~meV). We found it to be sufficient to fix this to [-400 meV, 400 meV].
    \item Since the energy window is much larger than the interval of interest, it is beneficial to use a non-uniform mesh that is denser near zero and sparser at higher energies. We use a triqs\_maxent's {\tt{HyperbolicOmegaMesh}}. We find the results were converged with a mesh of 750 points, which corresponds to a spacing of just under 0.4 meV at $10$~meV).
    \item The maxent procedure takes the Monte Carlo noise level as a parameter. We used a single (scalar) error value at all frequencies, and varied it as a free parameter. Since we chose the optimal value of the maxent hyperparameter using the line fit method~\cite{triqs_maxent}, this parameter is not expected to have a major effect on the results. By trial and error, we fixed this to be $10^{-1}$~meV.
    \item We choose the mesh for the hyperparameter $\alpha$ by manually inspecting the $\chi(\alpha)$ curve, such that the optimal $\alpha$ value was in near the center of the range, and there were enough points to find the optimal value of $\alpha$ with a line fit. We found that using 30 points in the interval $[10^{-9},10^{-2}]$ worked well.
    \item We found very little qualitative difference between the resulting self-energies with and without preblur. Therefore, we chose to do our calculations without preblur.
\end{itemize}

\section{Calculation of conductivity}
As stated in the main text, we calculate the conductivity as \cite{georges1996dynamical}
\begin{align}
	\sigma^\mathrm{dc}_{\alpha \beta} =  \frac{\pi e^2}{ \hbar S N }  \sum_{\sigma \bm{k}}  \int d\omega \Big(-\frac{\partial f}{\partial \omega} \Big) \mathrm{Tr}[ \bm{v}^\alpha_{\sigma \bm{k}} \bm{A}_{\sigma \bm{k}}(\omega) \bm{v}^\beta_{\sigma \bm{k}} \bm{A}_{\sigma \bm{k}}(\omega) ],
\end{align}
where $S$ is the area of the unit cell, $N$ the number of $\bm{k}$-points in the first Brillouin zone, and $f$ the Fermi-Dirac function. $\bm{v}^\alpha_{\sigma \bm{k}}$ is the Fermi velocity in direction $\alpha \in \{x, y\}$ evaluated in the orbital basis and supplemented with an additional term to correctly incorporate the Peierls substitution \cite{tomczak2009}. The matrix elements of $\bm{v}^\alpha_{\sigma \bm{k}}$ are given by
\begin{align}
[v^\alpha_{\sigma \bm{k}}]_{\eta \eta'} = \frac{1}{\hbar} \frac{\partial}{\partial{k_\alpha}}  [t_{\sigma\bm{k}}]_{\eta \eta'} - \frac{i}{\hbar}(\rho^\alpha_\eta -  \rho^\alpha_{\eta'}) [t_{\sigma\bm{k}}]_{\eta \eta'}.
\end{align}
Here, $t_{\sigma, i\eta, j\eta'}$ represents the hopping amplitude between orbital $\eta$ located in Bravais lattice cell $i$ and orbital $\eta'$ at cell $j$, for spin $\sigma$ character. $[t_{\sigma\bm{k}}]_{\eta \eta'} = \sum_{ij}t_{\sigma, i\eta, j\eta'} e^{-\bm{k} \cdot (\bm{R}_i - \bm{R}_j)}$ is the Fourier transform of $t_{\sigma, i\eta, j\eta'}$ with 
$\bm{R}_i$ the position of the cell $i$.  $\bm{\rho}^\alpha_\eta$ denotes the $\alpha$-component of the position vector of orbital $\eta$ within the unit cell. The first term corresponds to the conventional Fermi velocity, while the second term accounts for hopping amplitudes between different orbitals located at distinct sites within the unit cell \cite{tomczak2009}.
$\bm{A}_{\sigma \bm{k}}(\omega)$ is the momentum-resolved spectral function in the orbital basis whose matrix elements read
\begin{align}
[A_{\sigma \bm{k}}(\omega)]_{\eta \eta'} = \big[ [ (\omega + \mu) \bm{I}  - \bm{t}_{\sigma \bm{k}} - \bm{\Sigma}(\omega)]^{-1} \big]_{\eta \eta'},
\end{align}
where $\bm{I}$ is the identity matrix and $\mu$ the chemical potential.  $\bm{\Sigma}(\omega)$ is a diagonal self-energy matrix whose elements correspond to the DMFT self-energies of the T, H$_1$, and H$_2$ orbitals. 
It is obtained via analytic continuation of the self-energy on the Matsubara frequency axis to the real frequency $\omega$ axis.

\section{Single-orbital DMFT calculations of the topmost band}

\begin{figure} [!htbp]	 
	\includegraphics[width=0.9\columnwidth]{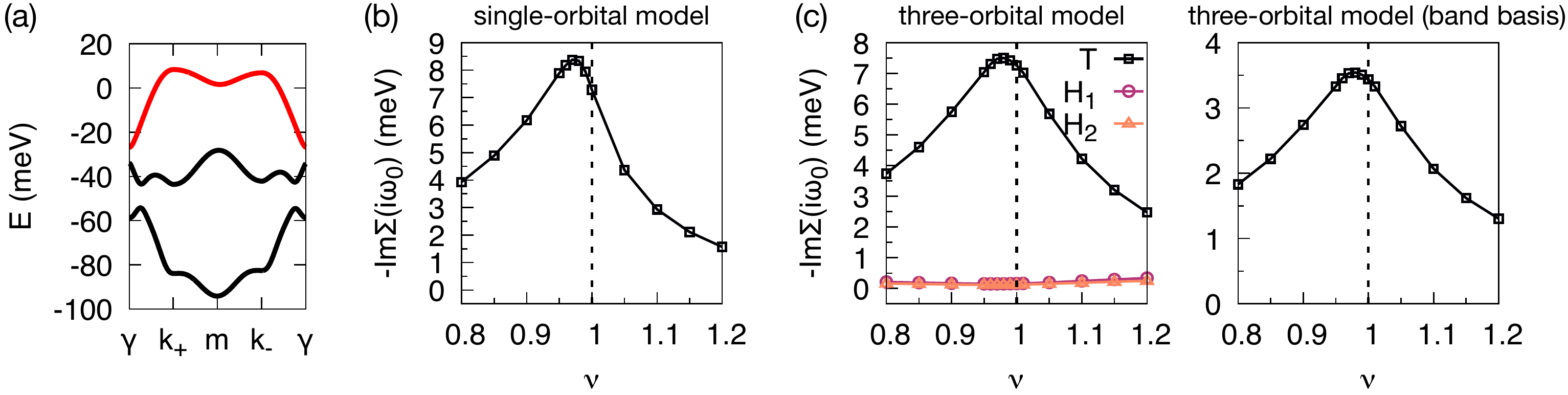}
	\caption{(a) The noninteracting band dispersion $\epsilon^\uparrow_{n}(\bm{k})$ at $E_z = 1$~meV. The topmost band ($n = 1$) is highlighted with a red line. (b) The calculated imaginary part of the DMFT self-energy at the lowest Matsubara frequency, obtained by solving Eq.~(\ref{singleH}) using DMFT at $U \simeq 0.9W$ ($W$: bandwidth of the topmost band), $E_z = 1$~meV, and $T = 5.8$~K. (c) The corresponding DMFT data from the original three-orbital model at $\varepsilon = 9.3$, $E_z = 1$~meV, and $T = 5.8$~K. Left: $-\mathrm{Im}\Sigma_{\eta}(i\omega_0)$ for orbital $\eta$ ($\eta \in \{ \mathrm{T}, \mathrm{H}_1, \mathrm{H}_2 \}$). Right: $-\mathrm{Im}\Sigma(i\omega_0)$ of the topmost band, defined as $-\mathrm{Im}\Sigma(i\omega_0) \equiv \frac{1}{N}\sum_{\bm{k} \eta}[P(\bm{k})]^{-1}_{n \eta} [-\mathrm{Im}\Sigma_{\eta}(i\omega_0)]P(\bm{k})_{\eta n } \big|_{n=1} $, where $P(\bm{k})_{\eta n}$ are elements of the basis transformation matrix from the orbital $\eta$ to the band $n$ at crystal momentum $\bm{k}$.   }
	\label{sfig_singleorb}
\end{figure}

To disentangle the effects of charge transfer and the vHS, we solve a single-orbital model constructed from the more general three-orbital Hamiltonian $H_0$. We first diagonalize $H_0$ and evaluate the band energy $\epsilon^\sigma_{n}(\bm{k})$ (with $n \in \{ 1,2,3 \}$ indicating the band index numbered from top to bottom); see Fig.~\ref{sfig_singleorb}(a). Then we Fourier transform only the topmost band ($n=1$) to the real space as
\begin{align}
    \tilde{t}_{\sigma, i j} = \frac{1}{N} \sum_{\bm{k}} \epsilon^\sigma_{n}(\bm{k}) e^{ \bm{k} \cdot (\bm{R}_i - \bm{R}_j)}.
\end{align}
Here, $\tilde{t}_{\sigma, ij}$ is the resulting hopping amplitude on the triangular lattice, which, together with the local interaction term, defines the following single-orbital Hamiltonian $\tilde{\mathcal{H}}$:
\begin{align} \label{singleH}
    \tilde{\mathcal{H}} = \sum_{ij, \sigma}\tilde{t}_{\sigma, i j} c^{\dagger}_{i\sigma} c_{j\sigma} - \mu \sum_{i, \sigma}\tilde{n}_{i\sigma} +  U\sum_{i}  \tilde{n}_{i\uparrow}\tilde{n}_{i\downarrow} ,
\end{align}
where $c^\dagger_{i\sigma}$  ($c_{i\sigma}$) is the electron creation (annihilation) operator for site $i$ and spin $\sigma$, $\tilde{n}_{i\sigma} = c^\dagger_{i\sigma} c_{i\sigma}$ the electron number operator, and $\mu$ the chemical potential which will be adjusted to achieve the desired filling during the DMFT self-consistency. $U$ denotes the local Coulomb interaction, which is an adjustable parameter in the model. Note that we neglect any nonlocal density-density interactions here because their Hartree treatment merely shifts the on-site energy level, which can be simply absorbed into the chemical potential.
By solving Eq.~(\ref{singleH}), we can clarify whether the scattering-rate asymmetry observed in the small $E_z$ regime (where no Mott transition occurs at $\nu = 1$) in the more general three-orbital model originates from the charge-transfer physics or from the vHS, since Eq.~(\ref{singleH}) includes only the latter effect.

Figure~\ref{sfig_singleorb}(b) presents the imaginary part of the DMFT self-energy at the lowest Matsubara frequency obtained by solving Eq.~(\ref{singleH}). Here, we clearly observe the same asymmetry between the hole- and electron-doped sides as in the T-orbital self-energy of the  original three-orbital model [cf. Fig.~\ref{sfig_singleorb}(c)]. It suggests that the asymmetry in the resistivity for the small $E_z$ regime [Fig.~3(a) in the main text] is attributed to the effect of the vHS rather than to the multiorbital charge-transfer physics.


%